\documentclass[12pt]{article}

\newcommand{\bbr}{I\!\! R}

\newcommand{\x}{arXiv:}
\newcommand{\m}{\mathrm}
\newcommand{\be}{\begin{equation}}
\newcommand{\ee}{\end{equation}}
\newcommand{\ba}{\begin{eqnarray}}
\newcommand{\ea}{\end{eqnarray}}

\usepackage{graphicx}
\usepackage{amssymb}
\usepackage{amsmath}

\begin{document}
\thispagestyle{empty}
\begin{center}

\null \vskip-1truecm \vskip2truecm

{\Large{\bf \textsf{A Holographic Bound on Cosmic Magnetic Fields}}}

{\large{\bf \textsf{}}}

{\large{\bf \textsf{}}}

\vskip1truecm

{\large \textsf{Brett McInnes}}

\vskip1truecm

\textsf{\\  National
  University of Singapore}

\textsf{email: matmcinn@nus.edu.sg}\\

\end{center}
\vskip1truecm \centerline{\textsf{ABSTRACT}} \baselineskip=15pt
\medskip

Magnetic fields large enough to be observable are ubiquitous in astrophysics, even at extremely large length scales. This has led to the suggestion that such fields are seeded at very early (inflationary) times, and subsequently amplified by various processes involving, for example, dynamo effects. Many such mechanisms give rise to extremely large magnetic fields at the end of inflationary reheating, and therefore also during the quark-gluon plasma epoch of the early universe. Such plasmas have a well-known holographic description in terms of a thermal asymptotically AdS black hole. We show that holography imposes an upper bound on the intensity of magnetic fields ($\approx \; 3.6 \times 10^{18}\;\; \text{gauss}$ at the hadronization temperature) in these circumstances; this is above, but not far above, the values expected in some models of cosmic magnetogenesis.

\newpage
\addtocounter{section}{1}
\section* {\large{\textsf{1. The Importance of Magnetic Fields in Cosmology}}}
One of the most pressing issues in astrophysics is the question of the origin of large-scale magnetic fields \cite{kn:reviewA,kn:reviewB,kn:kerstin}. These fields have been observed, using radiation at gamma-ray and other wavelengths, in a great variety of locales, including in intergalactic space \cite{kn:fermi}. This seems to find its most natural interpretation in terms of the hypothesis that magnetic fields are ``seeded'' by quantum fluctuations during inflation: that is, that the observed fields are \emph{cosmic} in origin. A fully satisfactory theory of cosmic magnetism remains, however, to be completed.

The importance of settling this question can hardly be over-stated. To take but two examples: the existence of cosmic magnetic fields may have a profound effect on the interpretation of recent claims that primordial gravitational B-modes have been observed in the cosmic microwave background \cite{kn:bicep}; and such fields may help to explain the reionization of the intergalactic medium \cite{kn:siri}.

There are however some serious difficulties facing this idea, of which the following is perhaps the most severe. Because Maxwell's equations in four dimensions are conformally invariant, and because all FRW spacetimes are conformally flat, one can construct an extremely general and robust argument to the effect that magnetic fields must decay adiabatically, that is, quite rapidly, with cosmic expansion. This makes it very difficult to obtain magnetic fields, at the present time, of the observed magnitude\footnote{Other approaches \cite{kn:reviewA} attempt to place the origin of the magnetic ``seed'' in the post-inflationary era, but these face equally or more serious problems with obtaining realistic coherence lengths for the field.}.

There are various ways of attacking this problem. One could explicitly break the conformal invariance of Maxwell's equations by, for example, positing an explicit non-minimal coupling to gravity: one can then try to use the variable coupling to moderate the dilution of the magnetic field \cite{kn:bamba}. Another ingenious approach \cite{kn:barrow1,kn:barrow2,kn:barrow3} circumvents the ``conformal'' argument by  exploiting the fact that marginally open FRW spacetimes are conformally flat in a slightly different sense to the spatially flat case. It is argued that, as a consequence, fluctuations of the magnetic field on scales larger than the spatial curvature scale (``supercurvature modes'') may lead to an anomalously slow decay (``superadiabatic amplification'') of the field. This slow decay might indeed be sufficient to solve the problems discussed earlier. If this were correct, then the existence of cosmic magnetism might be interpreted as direct evidence that the spatial sections of the Universe are negatively curved, a remarkable conclusion indeed.

Unfortunately, there are several serious objections to this claim: it seems that it may not be possible actually to excite supercurvature modes \cite{kn:durrer}, and that, even if it is possible, such modes do not in practice give rise to a significant amount of superadiabatic amplification of magnetic fields \cite{kn:sahni}. However, the question remains controversial \cite{kn:cost}.

More conservative explanations, that is, ones that accept the conventional evolution of the field after inflationary reheating, have been proposed: for example, a small magnetic field surviving through the inflationary era might be enormously amplified by dynamo-like effects during reheating ---$\,$ though a complete theory of such a dynamo remains to be constructed. (See the discussion around Figure 16 in \cite{kn:reviewB}.) In short, the claim here is that magnetic fields are observable at the present time not because they decay in some anomalous way, but simply because they were so large at the end of reheating. This hypothesis will be the main topic of this work.

We begin with the observation that this approach entails the existence of enormous magnetic fields during the \emph{plasma epoch} of cosmology. One might wonder whether such extreme\footnote{Following \cite{kn:reviewB}, we take ``extreme'' to mean that the energy density of the field is comparable to that of the plasma, a situation described as ``equipartition''. Under the usual assumptions, the ratio of these two densities remains constant during the plasma epoch, so we can apply the term consistently.} fields can really be sustained. Indeed, problems have been pointed out \cite{kn:afshordi} with specific models, but what we seek here is a \emph{general} upper bound on magnetic fields in such plasmas ---$\,$ general in the sense of being derived from some basic physical idea. We will argue here that some such bound must exist; the argument is based on \emph{holography}.

The basic observation here is that the plasma in question is a \emph{quark-gluon plasma} or QGP. Such plasmas hadronize when their temperatures fall sufficiently, but become very strongly coupled just prior to hadronization. The strongly-coupled QGP is notoriously difficult to handle theoretically, but several approaches have been developed. The one to be used here is known variously as ``gauge-gravity duality'' or simply as ``\emph{holography}''. This approach is based on the idea that certain \emph{strongly} coupled QCD-like field theories defined on $n\,-\;$dimensional spacetimes are physically equivalent (``dual'') to \emph{weakly} coupled gravitational theories in a spacetime of dimension $n+1$; these theories may incorporate extended objects, such as membranes (``branes'') propagating in the ``bulk''. By transferring the problem to this holographic dual theory, one can often gain new insights into unexpected features of the original field theory. An excellent semi-technical discussion of this approach to the strongly-coupled QGP is to be found in \cite{kn:mateos}.

To be more precise, the holographic description of the QGP \cite{kn:solana,kn:pedraza,kn:youngman,kn:gubser,kn:janik} reinterprets its properties in terms of those of a dual thermal black hole spacetime. Magnetic fields in the plasma correspond, in a way familiar from applications of holography to condensed matter physics, to a magnetic charge (per unit horizon area) on the black hole. Large values of that charge will give rise to a major deformation of the bulk spacetime, which in turn may have a strong effect on objects, such as branes, which inhabit the bulk and are sensitive to its geometry. It is not clear that such effects will always be benign, and we shall see that they are not. In this way, holography gives us a way of constraining the intensity of magnetic fields during the plasma epoch.

To be specific, we find that, if the magnetic field $B$ is sufficiently strong relative to the (squared) temperature, then the bulk black hole itself begins to generate branes, so that a static black hole picture is no longer consistent. The critical field strength is $2\,\pi^{3/2}T^2$; this turns out to be roughly an order of magnitude larger than the value required to generate the observed intergalactic magnetic fields. In view of the uncertainties attending holography generally, this can be interpreted to mean that the magnetic fields during the plasma epoch do satisfy the holographic bound, but only by a slim margin. One might speculate that holography somehow sets the scale of cosmic magnetism.

We begin by constructing a very simple holographic model of cosmic magnetic fields.

\addtocounter{section}{1}
\section* {\large{\textsf{2. Holography of Magnetic Fields in FRW Spacetimes}}}
Four-dimensional FRW spacetimes have an extremely restricted and simple geometry, which can be described in a variety of ways. Two of these ways are important here.

First, all FRW spacetimes are locally conformally flat; in the case where the spatial sections are flat, they are globally conformally flat (that is, the entire spacetime can be mapped to a flat spacetime by a single global conformal transformation; these are the FRW spacetimes of most interest, and we confine attention to them henceforth).

Secondly, all four-dimensional FRW spacetimes have the following property: at each point $p$ in three-dimensional space, each two-dimensional plane in the tangent space at $p$
can be mapped by a local isometry to any other such plane. This is just a way of formulating the condition that the space should be isotropic around every point (since, in three dimensions, there is a one-to-one correspondence between planes and their normal directions); but this way of stating the property is actually the more fundamental one. To see this, note that this condition, rather than isotropy \emph{per se}, is the one that allows us to deduce that the three-dimensional spacelike slices are spaces of constant curvature, this being a distinguishing feature of FRW cosmologies. This just means that planes in different \emph{locations} can also be mapped to each other isometrically\footnote{Here we are using Schur's (geometric) lemma (see \cite{kn:kobayashi}, page 202), which applies to any Riemannian manifold of dimension at least three. It states that, if the curvature evaluated on 2-dimensional planes (the sectional curvature) is independent, at each point, of the choice of plane, then it must also be independent of position, and hence be fully constant. Note that isotropy does \emph{not} imply constant curvature in dimensions higher than three, but pointwise independence of the choice of plane \emph{does} imply that conclusion in all such dimensions.}. In other words, FRW spacetimes can be thought of as the spacetimes in which \emph{spatial planes behave as simply as possible}: if one understands the physics associated with any one plane, then one understands the physics of all of them.

This point of view is particularly appropriate for discussing magnetic fields, because it is natural to associate a magnetic field with the corresponding \emph{flux} through a planar surface. In fact, it will be useful to take the point of view that the flux is fundamental, and the magnetic field is just a quantity deduced from the flux per unit area through some plane. This is appropriate from a holographic point of view because the flux (leaving aside superadiabatic amplification, which cannot occur in the spatially flat case) is actually \emph{conformally invariant}. To see this, notice that by solving the magnetic wave equation for the modes corresponding to a cosmic magnetic field $B$, one concludes \cite{kn:reviewA} that $B$ must decrease according to $a(t)^{-2}$, where $a(t)$ is the usual FRW scale factor; but then the flux through a plane\footnote{By ``plane'' here we of course really mean some compact domain in the plane. We take this as understood henceforth.} $\mathcal{S}$ remains constant, because the area of $\mathcal{S}$ increases according to $a(t)^2$.

All this suggests that we set up our holographic model of cosmic magnetism as follows. We take a FRW spacetime with flat spatial sections, containing a plasma and a magnetic flux associated with a fixed but arbitrary plane $\mathcal{S}$. A conformal transformation takes us to a flat spacetime, in which $\mathcal{S}$ retains its geometry but no longer evolves with time. Adjoining the time direction, we can use $\mathcal{S}$ to define a three-dimensional flat spacetime permeated by a magnetic field. We then study the holographic dual of this spacetime. The magnetic field and temperature in this spacetime are constant, but the time dependence can be restored when convenient by reverting the conformal transformation. In more detail: as we have just seen, the adiabatic dilution of $B$, to which we are adhering here, means that it decays according to $a(t)^{-2}$; on the other hand, the temperature of the plasma will decline according to $a(t)^{-1}$. Therefore the ratio $B/T^2$ is a conformal invariant in FRW geometry, and it can be evaluated in the conformally transformed boundary geometry; therefore it can be studied through the latter's holographic dual. This is our plan for using holography to study cosmic magnetism.

The plasma is a thermal system, so we need an asymptotically AdS black hole with a non-zero Hawking temperature in the bulk. The transverse sections (perpendicular to the radial direction, parallel to the event horizon of the black hole) will be copies of $\mathcal{S}$, so they must be planar. That is, we need a \emph{planar} black hole, of the kind that exists in the asymptotically AdS case \cite{kn:lemmo} because the negative cosmological constant violates the relevant energy condition.

We assume that the bulk black hole is ``dyonic'' (see for example \cite{kn:dyon}), that is, both electrically and magnetically charged; we then obtain the metric and electromagnetic potential from the conventional Einstein-Maxwell equations. The geometry is described by a ``Charged Planar AdS Black Hole'' metric, given by
\begin{eqnarray}\label{A}
g(\m{CPAdSBH}) & = & -\, \Bigg[{r^2\over L^2}\;-\;{8\pi M^*\over r}+{4\pi (Q^{*2}+P^{*2})\over r^2}\Bigg]dt^2\; \nonumber \\
& &  + \;{dr^2\over {r^2\over L^2}\;-\;{8\pi M^*\over r}+{4\pi (Q^{*2}+P^{*2})\over r^2}} \;+\;r^2\Big[d\psi^2\;+\;d\zeta^2\Big].
\end{eqnarray}
Here $\psi$ and $\zeta$ are dimensionless planar coordinates, $L$ is the asymptotic AdS curvature radius, and $M^*$, $Q^*$, and $P^*$ are geometric parameters with no direct physical meaning, but whose significance we now explain. In accordance with our emphasis on the role of planar geometry, we claim that the physical parameters for such a black hole are its \emph{mass per unit horizon area}, which we denote by $\mathcal{M}$, and the \emph{electric and magnetic charges per unit horizon area}, $\mathcal{Q}$ and $\mathcal{P}$. (In fact, the actual mass and charges of such a black hole are formally infinite, so it can only be described by using constructs of this kind.) $\mathcal{M}$, $\mathcal{Q}$, and $\mathcal{P}$ are related to $M^*$, $Q^*$, and $P^*$ as follows. First note that $M^*$, $Q^*$, and $P^*$ determine (for a fixed value of $L$) the value of $r$ at the event horizon, $r = r_h$; then we have simply
\begin{equation}\label{BETH}
\mathcal{M} \;=\;M^*/r_h^2,\;\;\;\;\mathcal{Q}\;=\;Q^*/r_h^2, \;\;\;\;\mathcal{P}\;=\;P^*/r_h^2;
\end{equation}
see \cite{kn:77} for a detailed discussion of similar formulae. Conversely, given $\mathcal{Q}$, $\mathcal{P}$, and the Hawking temperature of the black hole (see below), one can readily find $r_h$, and then use these relations to compute $M^*$, $Q^*$, and $P^*$ in terms of $\mathcal{M}$, $\mathcal{Q}$, and $\mathcal{P}$.

The potential form for the electromagnetic field outside the black hole is
\begin{equation}\label{B}
A\;=\;\left ({1\over r_h}\,-\,{1\over r}\right ){Q^*\over L}dt\;+\;{P^*\over L}\psi d\zeta,
\end{equation}
where the constant term in the coefficient of $dt$ is inserted so that the Euclidean version should be well-defined at the origin. The field strength form is
\begin{equation}\label{C}
F\;=\;-\,{Q^*\over r^2L}dt \wedge dr \;+\;{P^*\over L}d\psi \wedge d\zeta.
\end{equation}

The baryonic chemical potential $\mu_B$ of the dual system is related holographically to the asymptotic value of the time component of the potential form, that is, to $Q^*/(r_hL)$. In view of the value ($\eta_B \approx 10^{-9}$) of the net baryon density/entropy density ratio usually assumed to hold during the plasma era, which in turn is related to the ratio of the chemical potential to the temperature, we can take $\mu_B$ to be zero to an excellent approximation. We therefore set $Q^* = \mathcal{Q} = 0$ for the remainder of this work.

We see that, on the other hand, the magnetic field persists to infinity, thus opening the way to a holographic interpretation of magnetic fields on the boundary\footnote{As is well known, the U(1) on the boundary is actually global, but it can be gauged by considering the two different possible falloff behaviours for electromagnetic fields in asymptotically AdS spacetimes: see \cite{kn:klebwit}.}. This observation \cite{kn:hartkov} is familiar, and of fundamental importance, in applications of holography to condensed matter physics.

The metric at infinity, normalised so that $t$ represents proper time for a stationary observer, is such that the norms of $d\psi$ and $d\zeta$ are both equal to $1/L$, so the corresponding unit forms are $Ld\psi$ and $Ld\zeta$; thus, if $B$ is the magnetic field, we have, from equation (\ref{C}),
\begin{equation}\label{D}
B \;=\; P^*/L^3
\end{equation}
as the holographic relation between the magnetic parameter of the black hole (which determines its magnetic charge per unit horizon area in the manner explained earlier) and the magnetic field of the dual system on the boundary. (Notice that, in the units we use here, $P^*$ has the same units as $r$, so $B$ has units of inverse length squared, as it should.)

The temperature of this black hole can be found in the usual manner, by requiring that the Euclidean version of the geometry be regular: it is
\begin{equation}\label{E1}
T\;=\;{1\over 4\pi r_h}\,\Bigg({3r_h^2\over L^2}\;-\; {4\pi P^{*2} \over r_h^2}\Bigg),
\end{equation}
where we have used the definition of $r_h$, namely
\begin{equation}\label{E2}
{r_h^2\over L^2}\;-\;{8\pi M^*\over r_h}+{4\pi P^{*2}\over r_h^2} = 0,
\end{equation}
to eliminate the explicit dependence on $M^*$, which we do not need here.

The Hawking temperature in (\ref{E1}) will be interpreted holographically, in the usual manner, as the temperature of the plasma. Equations (\ref{D}),(\ref{E1}), and (\ref{E2}) allow us to translate between the geometric parameters $M^*$ and $P^*$ (which determine $r_h$, and subsequently fix the physical parameters $\mathcal{M}$ and $\mathcal{P}$) and their field-theory counterparts $T$ and $B$.

It is of interest to note that if we impose the very reasonable condition that $T \geq 0$ (which corresponds to requiring cosmic censorship on the black hole side of the duality), then we obtain an upper bound on the magnetic charge per unit horizon area, in terms of the asymptotic AdS curvature radius\footnote{It is remarkable that the condition for censorship to hold can be expressed exclusively in terms of $\mathcal{P}$, without comparing it with $\mathcal{M}$. Expressed in terms of the (less directly physical) parameter $P^*$, however, censorship does require a comparison with $M^*$, to wit, $P^{*6} \leq (27/4)\pi M^{*4}L^2$; this is obtained by requiring the quartic in (\ref{E2}) to have a positive real solution.}:
\begin{equation}\label{E3}
\mathcal{P} \;\leq \; \mathcal{P}^+\;=\;\sqrt{{3\over 4\pi}}\,{1\over L}.
\end{equation}
This will help us to understand how censorship violation is avoided in our subsequent work.

This black hole is the basis of our holographic study of cosmic magnetic fields. However, we observe immediately that equation (\ref{E1}) seems to indicate that one should \emph{not} expect to be able to establish any straightforward holographic relation between $T^2$ and $B$, as we are hoping to do in this work. In particular, the fact that both quantities have the same units is quite irrelevant, because there are two other parameters with units of length in the problem, namely $r_h$ and $L$; actually, because of this, (\ref{E1}) seems to relate $T$ to $B^2$ (see (\ref{D})) rather than $T^2$ to $B$. We will nevertheless see that holography can surmount these difficulties, in a remarkably elementary way.

In order to proceed, we must assume that this simple gravitational system is an approximate description of a full string-theoretic system in the bulk, with all of its attendant fields and objects, such as branes. The description will be a good one provided that the appropriate parameters (the string coupling, the ratio of the string length scale to the AdS curvature scale $L$) are sufficiently small, and provided that the additional objects can be consistently ignored. It can be difficult to ensure this last condition, as we shall discuss. To aid that discussion, we make a brief excursion into the geometry of the general class of spacetimes we are considering here, namely those which are asymptotically AdS and can be foliated by planar transverse sections.

\subsubsection*{{\textsf{3. Asymptotically AdS Spacetimes with Planar Transverse Sections}}}
As is well known, asymptotically AdS spacetimes, or submanifolds of them, can often be foliated in a variety of different ways. For example, a suitable submanifold of AdS$_4$ itself can be foliated by flat 3-dimensional subspaces transverse to a radial direction, so that the ``planar AdS'' metric takes the form
\begin{equation}\label{F}
g(\m{PAdS}) = -\, {r^2\over L^2}\,dt^2\; + \;{dr^2\over r^2/L^2} \;+\;r^2\Big[d\psi^2\;+\;d\zeta^2\Big];
\end{equation}
the coordinates here are as in equation (\ref{A}), from which (\ref{F}) is obtained by setting $M^* = P^* = 0$.

Now consider a transverse section (including time, as above) of the form $r = constant$. The ``volume'' form of such a section ---$\,$ for later convenience, let us think of it instead as an \emph{area} ---$\,$ will take the form
\begin{equation}\label{G}
\m{Area}_3 \;=\;{r^3\over L}dt \wedge d\psi \wedge d\zeta.
\end{equation}
Upon performing the integrals over a compact domain in the ($t, \psi, \zeta$) directions\footnote{Henceforth we shall assume that such a compact domain has been fixed, so that the areas and volumes of which we speak refer to that domain. This is just a formality to avoid infinite quantities. An alternative procedure is to work in the Euclidean domain: then $t$ is compactified, and it is natural to compactify also $\psi$ and $\zeta$. Note that, unlike electric charge, magnetic charge must \emph{not} be complexified in passing to the Euclidean domain, so actually the Euclidean and Lorentzian formulae are essentially identical in this particular case.}, we see that areas in this spacetime grow in the radial direction as $r^3$.

However, if we wished to compute the spacetime volume ``contained'' in this section --- we shall see later how to be more precise about this concept --- then we would consider
\begin{equation}\label{H}
\m{Volume}_4 \;=\;{r^2}dt \wedge d\psi \wedge d\zeta \wedge dr.
\end{equation}
Performing the integrals, we find that this volume is \emph{also} proportional to $r^3$. Thus we arrive at the conclusion that areas of planar transverse sections, and the corresponding volumes, grow at essentially the same rate towards infinity in spaces of constant negative curvature. This fact is of course well known in the case of transverse spherical surfaces.

When we consider spacetimes which have planar transverse sections but which are only \emph{asymptotically} AdS, the situation becomes more complicated. It is still true that areas and volumes grow at the same rate towards infinity \emph{at leading order} in an expansion in $r$; but this need not be true at higher orders. This gives us a useful way of expressing how far a given asymptotically AdS spacetime has been deformed away from pure AdS.

We can express this idea in a concrete way by defining, on any such (four-dimensional) spacetime with asymptotic curvature radius $L$, the following quantity. Let $A_r$ be the area of the ($t, \psi, \zeta$) surface $r = constant$, and let $V_r$ be the volume contained. Then we set
\begin{equation}\label{I}
\mathfrak{S}(r)\; \equiv \; A_r\;-\;{3\over L}\,V_r,
\end{equation}
with the understanding that this quantity is defined only up to an overall positive constant (which depends on the detailed choice of the domain of integration, and which we shall choose so that $\mathfrak{S}(r)$ is dimensionless).
The factor of $3/L$ is chosen partly for dimensional reasons, but mainly to ensure that the leading terms cancel, so that $\mathfrak{S}(r)$ does indeed probe the higher-order terms.

For the planar foliation of AdS$_4$ itself, we therefore have
\begin{equation}\label{J}
\mathfrak{S}_{PAdS}(r) \;=\; 0
\end{equation}
for all $r$ (where $r$ extends down to $r = 0$), so $\mathfrak{S}(r)$ probes the \emph{difference} between a given four-dimensional asymptotically AdS spacetime and AdS$_4$ itself. In other words, if we define AdS to be the spacetime in which areas and volumes associated with transverse planes are (by construction) the \emph{same}, then, in asymptotically AdS spacetimes foliated by planes such that $\mathfrak{S}(r)$ does not vanish, the interpretation is that either the area of a transverse section is larger than its volume, or the reverse. (Notice that $\mathfrak{S}(r)$ might be positive for some values of $r$, and negative for others, so both kinds of behaviour can arise within the same spacetime.)

For all black hole spacetimes, we define the ``volume'' discussed above as the volume \emph{outside} the event horizon, so that $r$ only extends down to $r = r_h$. This is reasonable, because the area of the section $r = r_h$ (which is not what is usually called the ``area of the event horizon'', since it includes the time direction) is zero, and so the volume should likewise vanish. (Thus, $\mathfrak{S}(r_h) = 0$ for all black holes.)

In the case of asymptotically AdS spacetimes foliated by transverse sections of the form $\bbr \times S^2$, such as the exterior AdS-Schwarzschild spacetime, we can define a quantity analogous to $\mathfrak{S}(r)$, in precisely the same way as above. It turns out that, for AdS black holes with such transverse sections, this quantity is always positive both near to the event horizon, and far away from it; in fact, it is probably \cite{kn:74} positive for all values of $r$, for all black holes with event horizons having spherical topology. That is, the area is larger than the volume in all those cases.

The same statement holds true in many cases when the transverse sections are planar. For example, consider the planar AdS black hole with neither electric nor magnetic charge (thus with metric obtained from equation (\ref{A}) by setting $P^* = 0$). A straightforward computation then yields\footnote{Here, for simplicity of presentation, we have taken $r \neq r_h = (8\pi M^*L^2)^{1/3}$. However, the correct value of $\mathfrak{S}_{PAdSBH}(r_h)$, namely zero, is obtained by taking the limit $r \rightarrow r_h$.}, up to an overall positive constant factor as always,
\begin{equation}\label{K}
\mathfrak{S}_{PAdSBH}(r)\;=\; \left (1 + \left (1 - {8\pi M^*L^2\over r^3}\right )^{-1/2}\right )^{-1},
\end{equation}
and it is clear that this is never negative.

One might wonder whether $\mathfrak{S}(r)$ can ever be negative, for \emph{any} black hole with a planar event horizon. As we shall now explain, this question actually has a deep physical meaning.

\subsubsection*{{\textsf{4. $\mathfrak{S}(r)$ as a Brane Action }}}
It turns out that $\mathfrak{S}(r)$ is of direct physical, as well as geometrical, significance. As was pointed out by Seiberg and Witten \cite{kn:seiberg} (see also the investigations of Witten and Yau \cite{kn:wittenyau}), this quantity is, up to a positive constant factor (related to the tension of the brane) precisely \emph{the action of a BPS 2-brane} wrapping the section $r = constant$. Since, for any black hole, $\mathfrak{S}(r_h) = 0$, we see that the question as to whether the area \emph{always} exceeds the volume now takes on considerable physical significance: for if the volume can outgrow the area for some range of values of $r$, this means that the action is lower in that region than it is near to the event horizon. A brane-antibrane pair nucleating in that region will therefore have no tendency to contract or tunnel back into the black hole, and so the system will be unable to remain static. In short, it will no longer be consistent to ignore, as in Section 2 above, the presence of specifically string-theoretic objects in the bulk: even if they are not present initially, they will become steadily more important as the black hole \emph{itself} generates them.

As we saw earlier, this is not a problem for the planar black hole spacetime when $P^* = 0$, that is, in the absence of a magnetic field on the boundary; and, as one might expect, we shall see that it is likewise not a problem when the magnetic field is small. But it is far from clear that there are no difficulties here when the field is \emph{large}, which, as we have seen, is certainly possible in the cosmological context. In short, it is quite possible that cosmic magnetic fields might correspond to a bulk geometry which is so distorted (away from the $P^* = 0$ case) that the black hole becomes subject to the instability we have just been discussing. It is clearly very important to establish whether this actually happens. We now proceed to establish the precise condition for that.

\begin{figure}[!h]
\centering
\includegraphics[width=0.55\textwidth]{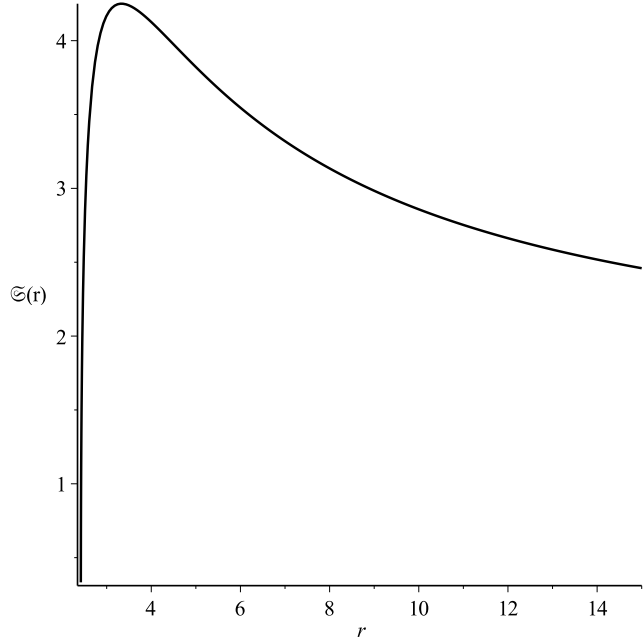}
\caption{$\mathfrak{S}_{CPAdSBH}(r)$ for $M^* = L = 1$, $P^{*6} = 3\pi$.}
\end{figure}

We can evaluate $\mathfrak{S}(r)$ for the metric in equation (\ref{A}): it is given (with $Q^* = 0$) by
\begin{equation}\label{L}
\mathfrak{S}_{CPAdSBH}(r)\;=\; {r^2\over L^2}\sqrt{{r^2\over L^2}-{8\pi M^*\over r}+{4\pi P^{*2}\over r^2}}-{r^3\over L^3}+{r_h^3\over L^3},
\end{equation}
which, after some simplifications, can be written as
\begin{equation}\label{M}
\mathfrak{S}_{CPAdSBH}(r)\;=\; { \left (-8\pi M^* + {4\pi P^{*2}\over r} \right )/L\over 1+\sqrt{1-{8\pi M^*L^2\over r^3}+{4\pi P^{*2}L^2 \over r^4}}}+{r_h^3\over L^3}.
\end{equation}
A typical graph of this function, for $P^*$ relatively small\footnote{For the parameter choices here and in Figure 2, see Footnote 10. The values for $M^*$ and $L$ were chosen for convenience, and then $P^*$ was selected so that censorship would be satisfied in both cases, and so that the graph would stay above the axis in one case, and cut it in the other.}, is shown in Figure 1.

One sees that $\mathfrak{S}_{CPAdSBH}(r)$ is zero at the event horizon, and positive near to it; however, after reaching a maximum it steadily decreases. It remains permanently positive for small $P^*$, but for larger $P^*$ it might eventually\footnote{It is clear from this description that the region of negative action, if it exists, will be located at large values of $r$, at least for values of the magnetic field not too far above some threshold; so one might wonder whether there is enough time for the instability to affect the entire geometry. That is not an issue in the present application, because the plasma epoch is extremely long-lived (several microseconds) by strong-interaction physics standards, so there is more than enough time for the instability to set in.} become negative; that can be avoided if and only if the limiting value as $r \rightarrow \infty$ is non-negative: that is, we need
\begin{equation}\label{N}
-\,4\pi M^* L^2 + r_h^3 \;\geq \;0.
\end{equation}
It is not obvious that this will always be satisfied, and in fact, for sufficiently large (but still sub-extremal) magnetic charge per unit horizon area, it is not: see Figure 2.

\begin{figure}[!h]
\centering
\includegraphics[width=0.55\textwidth]{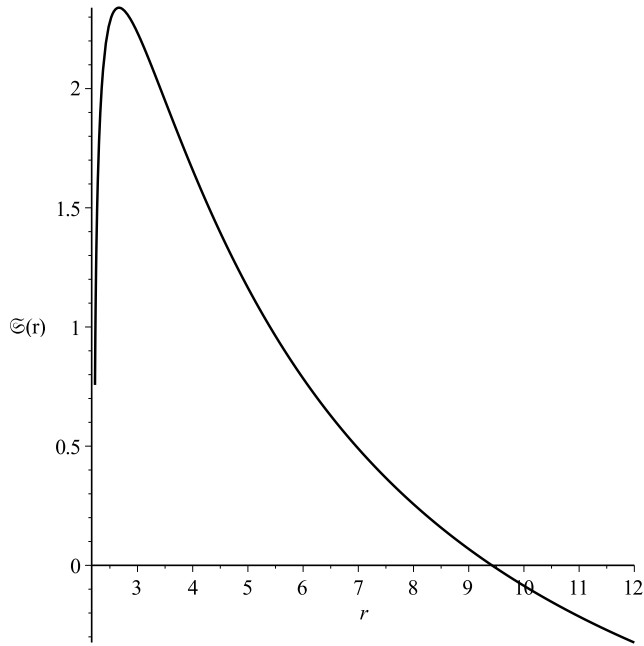}
\caption{$\mathfrak{S}_{CPAdSBH}(r)$ for $M^* = L = 1$, $P^{*6} = 5\pi$.}
\end{figure}

Thus, by requiring that the holographic dual of the system on the boundary should be well-behaved, we obtain a non-trivial restriction on the magnetic field. As we shall now show, that restriction is remarkably simple.

\subsubsection*{{\textsf{5. A Bound on the Magnetic Field }}}
Our objective is to turn the inequality (\ref{N}) into a relation involving $B$ and $T$, since those are the physical variables on the boundary. The difficulty is to do this while excluding $r_h$, which we do not want (and without resorting to solving a quartic for it, as in equation (\ref{E1}) or (\ref{E2})). With care, this can be done quite straightforwardly.

First we eliminate $M^*$ by using equation (\ref{E2}), and this converts (\ref{N}) to the form
\begin{equation}\label{O}
4\pi P^{*2}L^2 \;\leq \;r_h^4.
\end{equation}
Applying this to the second term on the right side of equation (\ref{E1}), we find
\begin{equation}\label{P}
2\pi TL^2 \geq r_h.
\end{equation}
Combining these two inequalities with equation (\ref{D}), \emph{we can actually eliminate both $r_h$ and $L$} and obtain an inequality involving $B$ and $T$ alone:
\begin{equation}\label{Q}
B\;\leq \;2\pi^{3/2}T^2\;\approx \; 11.14 \times T^2.
\end{equation}
This is the upper bound we seek.

Obviously this condition enforces a positive temperature if the magnetic field is non-zero; that is, it enforces cosmic censorship on the bulk side of the duality. To see how this works precisely, from the inequality (\ref{O}) one sees that the maximal magnetic charge per unit horizon area permitted if Seiberg-Witten instability is to be avoided, $\mathcal{P}^{SW}$, is given by
\begin{equation}\label{R}
\mathcal{P}^{SW}\;=\;{1\over \sqrt{4\pi}}\,{1\over L},
\end{equation}
which compares with the censorship value $\mathcal{P}^+$ (see inequality (\ref{E3})) according to
\begin{equation}\label{S}
{\mathcal{P}^{SW}\over \mathcal{P}^+} \;=\;{1\over \sqrt{3}} \;\approx \;0.577.
\end{equation}
Thus Seiberg-Witten instability sets in when the magnetic charge per unit horizon area is still below (though not far below) the value at extremality\footnote{If one prefers to work with the parameters $M^*$ and $P^*$, then the critical values are obtained by saturating inequalities (\ref{N}) and (\ref{O}), which yields $P^{*6} = 4\pi M^{*4}L^2$; this is in agreement with the censorship condition $P^{*6} \leq (27/4)\pi M^{*4}L^2$ obtained earlier. Note that Figures 1 and 2 were obtained by setting $P^{*6} = 3\pi M^{*4}L^2$ and $P^{*6} = 5\pi M^{*4}L^2$ respectively; both choices satisfy censorship, but the second clearly violates (\ref{Q}).}.

Let us consider some concrete values. A reasonable estimate for the hadronization temperature at the low values of the baryonic chemical potential we are considering here is $\approx$ 150 MeV. At that temperature, the inequality (\ref{Q}) takes the explicit form (in more familiar units)
\begin{equation}\label{T}
eB\;\leq \;\approx \; 3.6 \times 10^{18}\;\; \text{gauss}.
\end{equation}
This is just above the estimated maximum magnetic fields attained in peripheral heavy-ion collisions at the RHIC experiment \cite{kn:skokov}, which studies the QGP in precisely this region of the quark matter phase diagram\footnote{We should stress however that the plasma produced in heavy ion collisions is not directly comparable to the cosmic plasma; for example, while we are (as usual) neglecting vorticity in the cosmic plasma, the mechanism that generates large magnetic fields in heavy-ion plasmas also inevitably generates an enormous angular momentum density, which must not be neglected in a holographic description \cite{kn:79}. We are mentioning these experiments \emph{not} because we wish to claim that our bound applies to them, but merely in order to point out that magnetic fields on the order of $10^{18}$ gauss are by no means outlandish or unprecedented at these temperatures.}. (The ALICE experiment at the LHC studies the QGP at even higher values of the magnetic field, but also at considerably higher temperatures, so the ratio $B/T^2$ is similar.)

One can now consider whether a given theory of magnetogenesis is compatible with this bound. We begin with the main class of theory under consideration in this work, namely those in which the field decays adiabatically throughout the plasma era. In some such theories (see \cite{kn:reviewB}, Figure 16) the magnetic field satisfies equipartition at the end of reheating, that is, the energy density of the field is of the same order as the energy density of the plasma. The Stefan-Boltzmann law allows us to compute $B/T^2$ in that case; as we have discussed earlier, the adiabatic dilution of $B$ means that $B/T^2$ is constant (``conformally invariant''), so our bound will be satisfied at all times during the plasma era if it is satisfied at any one time. We can therefore check it at the point when the plasma hadronizes (as above, around $T = 150$ MeV). From the Stefan-Boltzmann law, we have under these circumstances
\begin{equation}\label{U}
B\;\approx\; \sqrt{{2\over 15}}\pi T^2 \;\approx\;1.15 \times T^2,
\end{equation}
or, in gauss,
\begin{equation}\label{UU}
eB\;\approx\; 3.7 \times 10^{17}\;\;\text{gauss}.
\end{equation}
We see that magnetic fields in these theories do satisfy the holographic bound, though they come within an order of magnitude of it.

A rival class of magnetogenesis scenarios \cite{kn:barrow1,kn:barrow2,kn:barrow3,kn:cost} holds that $B$ decays significantly more slowly than adiabatically during the plasma era, perhaps as slowly as $a(t)^{-1}$, where, as before, $a(t)$ is the cosmic scale factor. In that case, $B/T^2$ would, instead of remaining constant, actually \emph{grow} at the same rate as $a(t)$. During the plasma era, $a(t)$ grows by a very large factor, given by the ratio of the temperature of reheating to the temperature at hadronization; this ratio could be as large as $10^{17}$ if, for example, baryogenesis occurs at the grand unification scale. Thus even if $B/T^2$ is very small at reheating, one might well find that the holographic bound (\ref{Q}) is violated by the end of the plasma era; in other words, one has a ``fine-tuning'' problem of a degree of severity that depends on one's theory of reheating. However, questions of initial conditions, as discussed extensively in \cite{kn:cost}, are subtle in this kind of theory, so we do not assert that such an approach is definitely in conflict with holography; but it does seem clear that holography generally favours the more conventional static evolution law for $B/T^2$.

Returning, then, to the case in which $B$ does dilute adiabatically: we saw that such theories lead to a magnetic field at hadronization which is about an order of magnitude below the holographic bound. In view of the many uncertainties arising in such applications of holography \cite{kn:mateos}, we prefer to state the case more tentatively: there is very likely to be a holographic bound on cosmic magnetic fields, and the fields actually occurring during the plasma epoch of the early Universe \emph{may well come close to reaching that bound}.

\subsubsection*{{\textsf{6. Conclusion: The Uses of Holography in Cosmology}}}
We have seen that cosmic magnetic fields can be usefully constrained by holography, even when we use the simplest possible structure for the bulk: we were able to find an upper bound on the magnetic field in the plasma epoch, in terms of a definite multiple of the squared temperature. Much more sophisticated techniques have of course been developed in holographic studies of magnetic fields in condensed matter contexts \cite{kn:albjohn}, and it would be interesting to adapt some of those methods to the cosmic case.

Even with the simplest model, the results are suggestive. It is remarkable that the holographic estimate of the largest possible field is so close to the values needed to obtain sufficiently large fields at the present time. Perhaps a deeper investigation will reveal more concrete evidence that holography not only constrains, but actually determines, the scale of magnetic fields during the plasma epoch of the early Universe. It might be useful to pursue this idea in the context of string-theoretic attempts to account for the seeding of cosmic magnetic fields; see for example \cite{kn:rhiannon}.

\addtocounter{section}{1}
\section*{\large{\textsf{Acknowledgement}}}
The author is grateful to Dr. Soon Wanmei for helpful discussions.


\begin{thebibliography}{18}
\bibitem{kn:reviewA}
Alejandra Kandus, Kerstin E. Kunze, Christos G. Tsagas, Primordial magnetogenesis, Phys. Rep. 505 (2011) 1, arXiv:1007.3891 [astro-ph.CO]
\bibitem{kn:reviewB}
Ruth Durrer, Andrii Neronov, Cosmological Magnetic Fields: Their Generation, Evolution and Observation, Astron.Astrophys.Rev. 21 (2013) 62, arXiv:1303.7121 [astro-ph.CO]
\bibitem{kn:kerstin}
Kerstin E. Kunze, Cosmological Magnetic Fields,	Plasma Phys.Control.Fusion 55 (2013) 124026, arXiv:1307.2153 [astro-ph.CO]
\bibitem{kn:fermi}
Ievgen Vovk, Andrew M. Taylor, Dmitri Semikoz, Andrii Neronov, Fermi/LAT observations of 1ES 0229+200: implications for extragalactic magnetic fields and background light, Ap. J. Lett. 747 (2012) L14, arXiv:1112.2534 [astro-ph.CO]
\bibitem{kn:bicep}
Camille Bonvin, Ruth Durrer, Roy Maartens, Can primordial magnetic fields be the origin of the BICEP2 data?, Phys.Rev.Lett. 112 (2014) 191303, arXiv:1403.6768 [astro-ph.CO]
\bibitem{kn:siri}
Sirichai Chongchitnan, Avery Meiksin, The Effect of Cosmic Magnetic Fields on the Metagalactic Ionization Background inferred from the Lyman-alpha Forest,
Mon.Not.Roy.Astron.Soc. 437 (2014) 3639, arXiv:1311.1504 [astro-ph.CO]
\bibitem{kn:bamba}
Kazuharu Bamba, Sergei D. Odintsov, Inflation and late-time cosmic acceleration in non-minimal Maxwell-F(R) gravity and the generation of large-scale magnetic fields, JCAP 0804 (2008) 024, arXiv:0801.0954 [astro-ph]
\bibitem{kn:barrow1}
John D. Barrow, Christos G. Tsagas,
Cosmological magnetic field survival, Mon.Not.Roy.Astron.Soc. 414 (2011) 512, arXiv:1101.2390 [astro-ph.CO]
\bibitem{kn:barrow2}
J.D. Barrow, C.G. Tsagas, K. Yamamoto, 	
Origin of cosmic magnetic fields: Superadiabatically amplified modes in open Friedmann universes, Phys.Rev. D86 (2012) 023533, arXiv:1205.6662 [gr-qc]
\bibitem{kn:barrow3}
J. D. Barrow, C. G. Tsagas, K. Yamamoto,
Do intergalactic magnetic fields imply an open universe?
Phys.Rev.D86:107302,2012, arXiv:1210.1183 [gr-qc]
\bibitem{kn:durrer}
Julian Adamek, Claudia de Rham, Ruth Durrer, Mode Spectrum of the Electromagnetic Field in Open Universe Models,	Mon.Not.R.Astron.Soc. 423 (2012) 2705, arXiv:1110.2019 [gr-qc]
\bibitem{kn:sahni}
Yuri Shtanov, Varun Sahni, Can a marginally open universe amplify magnetic fields?,	JCAP 01 (2013) 008, arXiv:1211.2168 [astro-ph.CO]
\bibitem{kn:cost}
Christos G. Tsagas,	On the magnetic evolution in Friedmann universes and the question of cosmic magnetogenesis, arXiv:1412.4806 [astro-ph.CO]
\bibitem{kn:afshordi}
Takeshi Kobayashi, Niayesh Afshordi, Schwinger Effect in 4D de Sitter Space and Constraints on Magnetogenesis in the Early Universe,  JHEP 1410 (2014) 166,  arXiv:1408.4141 [hep-th]
\bibitem{kn:mateos}
David Mateos,
Gauge/string duality applied to heavy ion collisions: Limitations, insights and prospects, J.Phys.G G38 (2011) 124030, arXiv:1106.3295 [hep-th]
\bibitem{kn:solana}
Jorge Casalderrey-Solana, Hong Liu, David Mateos, Krishna Rajagopal, Urs Achim Wiedemann,
Gauge/String Duality, Hot QCD and Heavy Ion Collisions, arXiv:1101.0618 [hep-th]
\bibitem{kn:pedraza}
Mariano Chernicoff, J. Antonio Garcia, Alberto Guijosa, Juan F. Pedraza,
Holographic Lessons for Quark Dynamics, J.Phys.G G39 (2012) 054002, arXiv:1111.0872 [hep-th]
\bibitem{kn:youngman}
Youngman Kim, Ik Jae Shin, Takuya Tsukioka, Holographic QCD: Past, Present, and Future, Prog.Part.Nucl.Phys. 68 (2013) 55, arXiv:1205.4852 [hep-ph]
\bibitem{kn:gubser}
Oliver DeWolfe, Steven S. Gubser, Christopher Rosen, Derek Teaney, Heavy ions and string theory,  Prog.Part.Nucl.Phys. 75 (2014) 86, arXiv:1304.7794 [hep-th]
\bibitem{kn:janik}
Romuald A. Janik, AdS/CFT and applications,  PoS EPS-HEP2013 (2013) 141, arXiv:1311.3966 [hep-ph]
\bibitem{kn:kobayashi}
S. Kobayashi, K. Nomizu {\em Foundations of Differential Geometry I}, Interscience, 1963
\bibitem{kn:lemmo}
J.P.S. Lemos, Phys.Lett.B353:46,1995,
\x gr-qc/9404041; R.B. Mann, Class.Quant.Grav. 14 (1997) L109, arXiv:gr-qc/9607071;
Rong-Gen Cai, Yuan-Zhong Zhang, Phys.Rev.D54:4891,1996, \x gr-qc/9609065;
Danny Birmingham, Class.Quant.Grav. 16 (1999) 1197, arXiv:hep-th/9808032
\bibitem{kn:dyon}
Marco M. Caldarelli, Oscar J.C. Dias, Dietmar Klemm, Dyonic AdS black holes from magnetohydrodynamics, JHEP 0903 (2009) 025, arXiv:0812.0801 [hep-th]
\bibitem{kn:77}
Brett McInnes, Shearing Black Holes and Scans of the Quark Matter Phase Diagram, Class. Quantum Grav. 31 (2014) 025009, arXiv:1211.6835 [hep-th]
\bibitem{kn:klebwit}
Igor R. Klebanov, Edward Witten, AdS / CFT correspondence and symmetry breaking, Nucl.Phys. B556 (1999) 89, hep-th/9905104	 	
\bibitem{kn:hartkov}
Sean A. Hartnoll, Pavel Kovtun, Hall conductivity from dyonic black holes, Phys.Rev. D76 (2007) 066001, arXiv:0704.1160 [hep-th]
\bibitem{kn:74}
Brett McInnes, Kerr Black Holes Are Not Fragile,
Nucl. Phys. B 857 (2012) 362, arXiv:1108.6234 [hep-th]
\bibitem{kn:seiberg}
Nathan Seiberg, Edward Witten, The D1/D5 System And Singular CFT,
JHEP 9904 (1999) 017, \x hep-th/9903224
\bibitem{kn:wittenyau}
Edward Witten, Shing-Tung Yau,
Connectedness of the boundary in the AdS / CFT correspondence,Adv.Theor.Math.Phys.3:1635-1655,1999,
\x hep-th/9910245
\bibitem{kn:skokov}
V. Skokov, A.Yu. Illarionov, V. Toneev, Estimate of the magnetic field strength in heavy-ion collisions, Int.J.Mod.Phys. A24 (2009) 5925, arXiv:0907.1396 [nucl-th]
\bibitem{kn:79}
Brett McInnes, Angular Momentum in QGP Holography, Nucl. Phys. B887 (2014) 246, arXiv:1403.3258 [hep-th]
\bibitem{kn:albjohn}
Tameem Albash, Clifford V. Johnson, Scott MacDonald, Holography, Fractionalization and Magnetic Fields, Lect.Notes Phys. 871 (2013) 537, arXiv:1207.1677 [hep-th] \bibitem{kn:rhiannon}
Rhiannon Gwyn, Stephon H. Alexander, Robert H. Brandenberger, Keshav Dasgupta, Magnetic Fields from Heterotic Cosmic Strings, Phys.Rev. D79 (2009) 083502, arXiv:0811.1993 [hep-th]

\end{thebibliography}
\end{document}